\documentclass[%
rsi
]{revtex4-1}

\usepackage{graphicx}% Include figure files
\usepackage{dcolumn}% Align table columns on decimal point
\usepackage{bm}% bold math

\usepackage{geometry}                % See geometry.pdf to learn the layout options. There are lots.
\usepackage{graphicx}
\usepackage{amssymb}
\usepackage{epstopdf}
\usepackage{amsmath}
\usepackage{alltt}
\usepackage{wrapfig}
\usepackage{multirow}
\usepackage{rotating}
\usepackage{subfig}

\begin{document}

\title{ Monte Carlo Comparisons to a Cryogenic Dark Matter Search Detector with low Transition-Edge-Sensor Transition Temperature }

\author{
S.W.~Leman$^{a*}$, K.A.~McCarthy$^a$, P.L.~Brink$^b$, B.~Cabrera$^b$, M.~Cherry$^b$, E.~Do~Couto~E~Silva$^c$, E.~Figueroa-Feliciano$^a$, P. Kim$^c$, N.~Mirabolfathi$^d$, M.~Pyle$^b$, R.~Resch$^c$, B.~Sadoulet$^d$, B.~Serfass$^d$, K.M.~Sundqvist$^d$, and A.~Tomada$^b$, and B.A.~Young$^e$ 
}

\affiliation{
(a) MIT Kavli Institute for Astrophysics and Space Research, Cambridge, MA, U.S.A. \\
(b) Department of Physics, Stanford University, Stanford, CA, U.S.A. \\
(c) SLAC National Accelerator Laboratory, Menlo Park, CA, U.S.A. \\
(d) Department of Physics, The University of California at Berkeley, Berkeley, CA, U.S.A. \\
(e) Department of Physics, Santa Clara University, Santa Clara, CA, U.S.A. \\
* Corresponding Author's e-mail address: swleman@mit.edu \\
}

\begin{abstract}
We present results on phonon quasidiffusion and Transition Edge Sensor (TES) studies in a large, 3~inch diameter, 1~inch thick [100] high purity germanium crystal, cooled to 50~mK in the vacuum of a dilution refrigerator, and exposed with 59.5~keV gamma-rays from an Am-241 calibration source. We compare calibration data with results from a Monte Carlo which includes phonon quasidiffusion and the generation of phonons created by charge carriers as they are drifted across the detector by ionization readout channels. The phonon energy is then parsed into TES based phonon readout channels and input into a TES simulator. 
\end{abstract}

\maketitle

\section{Introduction}

The Cryogenic Dark Matter Search~\cite{Ahmed2009} utilizes silicon and germanium detectors to search for Weakly Interacting Massive Particle (WIMP) dark matter~\cite{Spergel2007,Tegmark2004} candidates. The silicon or germanium nuclei provide a target mass for WIMP-nucleon interactions. Simultaneous measurement of both phonon energy and ionization energy provide a powerful discriminator between electron-recoil (relatively high ionization) interactions and nuclear-recoil interactions (relatively low ionization). Background radiation primarily interacts through electron-recoils whereas a WIMP signal would interact through nuclear-recoils. The experiment is located in the Soudan Mine, MN, U.S.A.

The most recent phase of the CDMS experiment has involved fabrication, testing and commissioning of large germanium crystals and we present results on quasidiffusion studies in which the crystal was exposed with 59.5~keV gamma-rays from a collimated Am-241 calibration source. Prompt phonons are generated from electron-recoil interactions along with Neganov-Luke~\cite{Neganov1985, Luke1988} phonons created by charges as they drift through the crystal via the ionization channels' electric field. In our Monte Carlo, phonon transport is described by quasidiffusion~\cite{Marris1990, Tamura1993}, which includes anisotropic propagation, isotope scattering and anharmonic decay, until the phonons are absorbed in either the Transition Edge Sensor (TES)~\cite{Irwin1995} based phonon channels (photolithographically patterned on the flat crystal surfaces) or lost in surface interactions. The small fraction of surface covered by the TES surfaces and the low surface losses result in phonon pulse constants much longer than the TES time constants allowing for good observation of the underlying phonon physics.

%The phonon quasidiffusion measurements presented here differ from similar heat-pulse experiments in which the detector is partially submerged in a pumped liquid helium bath~\cite{Msall1997}. These crystals are situated in vacuum, which allows for reduced phonon losses at the surfaces. Additionally, these crystals were cooled to 50~mK in a dilution refrigerator. Phonon losses in the TESs and non-instrumented surfaces can be determined via the partitioning of energy in the phonon channels, and phonon propagation parameters are studied via signal timing in the phonon channels.

\section{CDMS Detectors}

The CDMS detectors are made of [001] high purity germanium crystals with [110] oriented flats. The CDMS detectors are 3~inches in diameter and 1~inch thick with a total mass of about 607~grams.

\begin{figure}[t]
\begin{tabular}{c}
\begin{minipage}{0.48\hsize}
\begin{center}
\includegraphics[width=7cm]{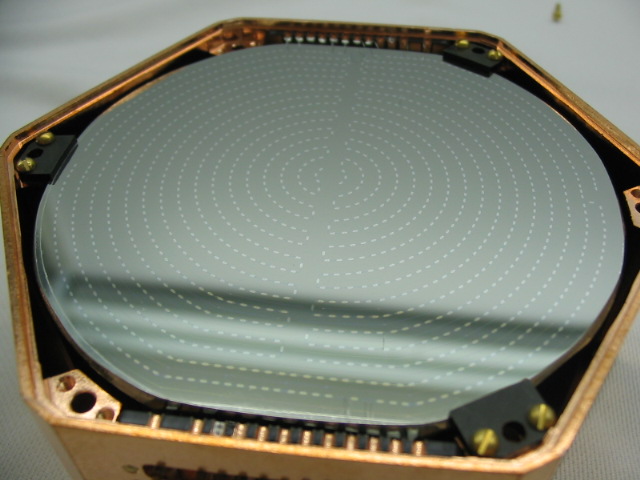}
\end{center}
\end{minipage}
\begin{minipage}{0.48\hsize}
\begin{center}
\includegraphics[width=7cm]{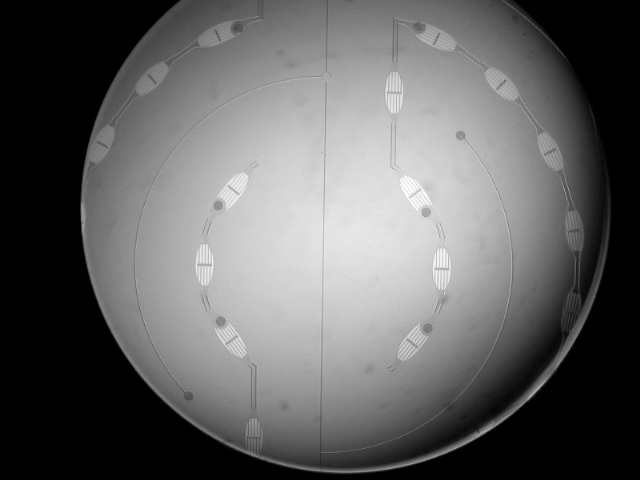}
\end{center}
\end{minipage}
\end{tabular}
\caption[] { \label{fig:iZip}(left) A CDMS version 2 iZIP detector with photolithographically defined phonon sensors. The crystal is 3~inches in diameter and mounted in its copper housing. The top surface contains an outer, guard phonon sensor and two inner phonon sensors from which an event's $x$ position estimate can be made. The opposite face (not shown) has a similar channel design, but rotated 90~degrees to determine an event's $y$ position estimate.}
\caption[] { \label{fig:iZipTES}(right) Close-up view of the iZIP phonon channel and ionization channel (thin lines in between the phonon sensors). The phonon channel is held at ground and the ionization channel is held at $\sim\pm$2~V for the top (bottom) surfaces.}
\end{figure}

The CDMS iZIP detector utilizes both anode and cathode lines on the same side of the detector similar to a Micro-Strip Gas Chamber (MSGC)~\cite{Oed1988, Luke1994, Brink2006} as shown in Figure~\ref{fig:iZip} and~\ref{fig:iZipTES}. Unlike an MSGC however, there is a set of anode and cathode lines on both sides of the detector. This ionization channel design is used to veto events interacting near the detector surfaces. The phonon sensor design described in this paper is for a version 2 iZIP and incorporates a guard ring channel, two channels on the top surface, from which $x$ position estimates can be made and two channels on the bottom surface, from which $y$ position estimates can be made. Versions that will be commissioned in Soudan have a modified phonon sensor design. The total iZIP aluminum coverage is 6.3\% per side, significantly reduced to previous versions of CDMS detectors.

Phonons are detected when they are absorbed into the aluminum fins, break Cooper pairs, and these Cooper pairs diffuse into the tungsten Transition Edge Sensors (TES). The TESs are superconductors, voltage biased within the superconducting to normal transition and the introduction of excited Cooper pairs raises their temperature which raises their resistance. Properly instrumenting the TESs allows the phonon flux power to be estimated.

\section{The CDMS Detector Monte Carlo, Phonons}

The CDMS detector Monte Carlo (CDMS-DMC)~\cite{Leman2011_5} models both the charge and phonon dynamics; these signals combined provide information about event location within the detector, the amount of energy deposited in the detector and the interaction recoil type (electron-recoil versus nuclear recoil). Partitioning of energy between the inner ionization channel and the outer, guard ionization channel combined with partitioning of energy between the four phonon channels provides information about event location. Event location provides a discriminator against surface event contamination, a class of event vetoed in low-background dark matter search operation and additionally allows for a position dependent energy calibration to be performed. The ratio of energy in the ionization channel vs. energy in the phonon channel indicates recoil type and provides a discriminator against gamma and beta background radiation, additional event types vetoed in low-background operation. If and when a population of WIMPs is detected, the spectrum of deposited energy will provide information regarding the WIMP mass~\cite{Yellin2002}. A proper modeling of the CDMS detectors is therefore necessary to understand and interpret the detector's signal and to properly veto background events.
	
In the CDMS-DMC, prompt phonons are produced at the interaction point with 2~THz energy and isotropic distribution of wave momentum and slow-transverse, fast-transverse and longitudinal mode density of states weighted by $\langle v_p \rangle ^{-3}$ , where $v_p$ is the phase velocity and brackets indicate an average. Their propagation is described by quasidiffusion, and the CDMS-DMC incorporates phonon focusing, isotope scattering and an isotropic approximation for anharmonic decay. 

At the detector surfaces, phonons are either reflect or absorbed. Phonons absorbed on non-instrumented walls are removed from the Monte Carlo. Phonon absorption on instrumented surfaces requires and energy of at least twice the aluminum superconducting gap in order to break a Cooper pair. The remainder of the phonon energy can either break additional Cooper pairs of escape back into the crystal as phonon energy. Phonons with energy below twice the aluminum superconducting gap cannot be readout in our phonon channels and are removed from the Monte Carlo.

\section{The CDMS Detector Monte Carlo, Charge}

\begin{figure}
\begin{center}
\includegraphics[width=12cm]{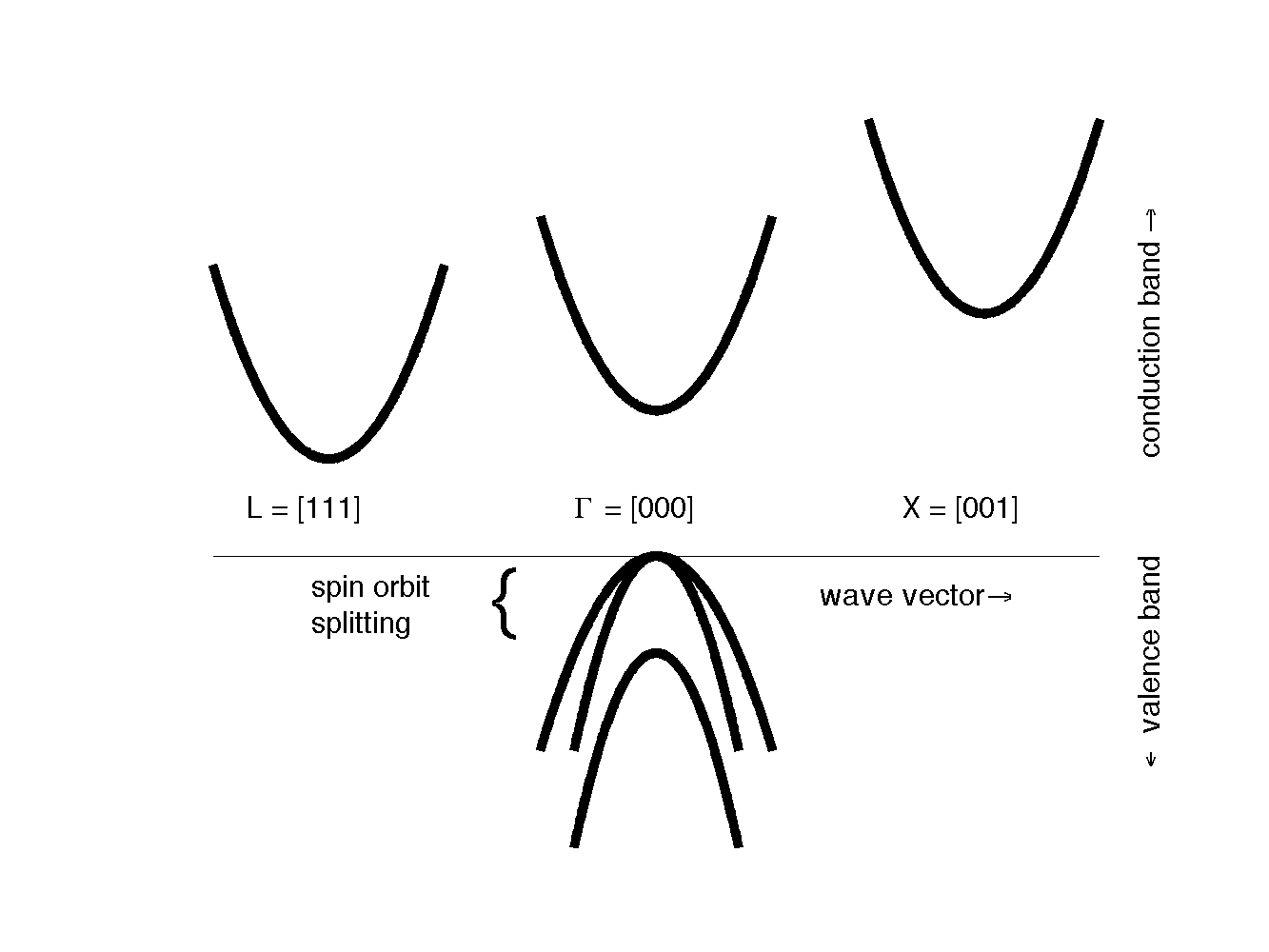}
\end{center}
\caption[] { \label{fig:bandGe} Germanium band structure showing the hole ground state, $\Gamma$ band and electron ground state L bands.}
\end{figure}

It is desirable to include a charge Monte Carlo for numerous reasons. First, the ionization signal, compared to the phonon signal, provides a discriminator between electron-recoil and nuclear-recoil events in the CDMS detectors. Electron transport is described by a mass tensor, leading to electron transport which is oblique to the applied field~\cite{Sasaki1958, Jacoboni1983} and is necessary to explain and interpret signals in the primary and guard-ring ionization channels~\cite{Cabrera}. Second, charges drifting through the detector produce a population of phonons, which contribute $\sim$25\% of the total phonon signal. These phonons' spatial, time, energy and emitted-direction distributions should therefore be properly modeled in the CDMS-DMC. Third, phonons created during electron-hole recombination at the surfaces contribute $\sim$25\% of the total phonon signal but in a low frequency, ballistic regime that is used to provide a surface-event discriminator. These phonons also need to be properly modeled in the CDMS-DMC. Germanium has an anisotropic band structure described schematically in Figure~\ref{fig:bandGe}. At low field, and low temperature, germanium's energy band structure in such that the hole ground state is situated in the $\Gamma$ band's [000] location and the electron ground state is in the L-band [111] location. Hole propagation dynamics are relatively simple due to propagation in the $\Gamma$ band and the isotropic energy-momentum dispersion relationship $\epsilon_{hole}(\mathbf{k}) = \hbar^2 k^2 / 2m$. Electron propagation dynamics are significantly more complicated due to the band structure and anisotropic energy-momentum relationship. At low fields and low temperatures, electrons are unable to reach sufficient energy to propagate in the $\Gamma$ or X-bands, and are not considered in the DMC. The electron energy-momentum dispersion relationship is anisotropic and given by $\epsilon_{electron}(\mathbf{k}) = \hbar^2 /2 (k_{\parallel}^2 / m_{\parallel} + k_{\perp}^2 / m_{\perp})$, where the longitudinal and transverse mass ratio $m_\parallel/m_\perp \sim$19.5. The SCDMS detectors are generally operated with low $\sim$1~V potential difference between the ionization channels and the phonon channels (which are held at ground). The potential accelerates the charge carriers towards the appropriate detector face. The holes propagate parallel to the field, in the $+z$ direction. The electrons however, due to the band anisotropy, propagate in the four L valleys and oblique to the electric field ($\sim$33 degrees from the $-z$ axis).
  
%\begin{figure}
%\begin{center}
%\includegraphics[width=12cm]{figures/129.png}
%\end{center}
%\caption[] { \label{fig:129} Gamma-ray event in the CDMS-DMC. Holes propagate upwards in the $\Gamma$ band. The electrons propagate oblique to the electric field in the four L bands. Charge carriers are shown as black points. Slow transverse, fast transverse and longitudinal phonons are shown as red, green and blue points, respectively.}
%\end{figure}

The charge carriers cannot accelerate indefinitely via the field and eventually shed phonons to limit their velocity to the longitudinal phase velocity. Modeling of these scattering processes is complicated by anisotropy and can be handled either via a Herring-Vogt~\cite{Herring1956, Cabrera} transformation, or rejection sampling methods~\cite{Sundqvist2009}. At these low fields, the electrons (holes) emit phonons with a peak frequency of $\sim$0.2 (0.3)~THz and are ballistic.

\section{The CDMS Detector Monte Carlo, Transition Edge Sensors}

Transition Edge Sensors are tungsten lines, voltage biased in their superconducting-normal transition allowing them to function as highly sensitive thermometers. The TESs readout the phonon energy absorbed in the aluminum fins. The TESs track the flux of phonons and therefore function like bolometers with a time constant $\tau_{TES} \sim C/G$, where $C$ is the heat capacity and $G$ is the thermal conductivity between the electron and phonon systems in the TES. The thermal recovery $\tau_{TES}$ is sped up by negative electrothermal feedback~\cite{Irwin1995} but while constant throughout the pulse is dependent on the TES's transition temperature $T_c$. For TESs with $T_c$ well above the substrate temperature the speed up is large, possibly and order of magnitude or more whereas for for TESs with $T_c$ near the substrate temperature, the speed up is of order 1. This causes the TES to be slow in response to phonon flux changes and affects the pulse shape.

TES channels on the top iZIP surface were modelled with a $T_c$ of 105~mK, consistent with constant current-flux measurements. The TESs on the bottom iZIP surface had low $Tc$, close to the substrate temperature which makes a precise $T_c$ measurement difficult. A direct measurement of $T_c$ with a constant current-flux method yielded a $T_c$ of 46~mK. A separate measurement can be made utilizing the TESs' current voltage (I-V) curves. When converted to current-power, the curves contain a constant power region as the TES goes through the 1~mK wide transition at $T_c$. From knowledge of the TES volume and electron-phonon coupling in the TES, this power can be converted to a temperature by equating the Joule heating and substrate cooling $P_{Joule} = P_{e \rightarrow \phi} = \Sigma \mathit{Vol} (T_e^n - T_\phi^n)$. For these channels the result was an 60~mK measurement of $T_c$, inconsistent with the former result.

The lack of knowledge of $T_c$ and its effect on pulse shape features require us to include a TES simulator in our model of the detector response. The Monte Carlo simulates the TES as 100 discrete elements per channel, extended across the surface. The phonon energy is parsed in position and input into the correct TES. Multiple simulations, utilizing 50, 55 and 60~mK variations of $T_c$ were run to determine a best fit between Monte Carlo and data.

\section{Experiments}

In the first round of experiments, a SCDMS mZip detector was exposed with an Am-241 gamma source to produce a population of prompt phonons and charge carriers within the detector. The detector ionization channels were biased at 6~V, producing a large population of Luke phonons. The detector was operated with the outer phonon channel segmented into three channels, which were individually instrumented and rotated 60~degrees relative to the inner three channels. 

The probability of a phonon being absorbed when it encounters a surface was calculated in another procedure and found to be 1.5\% for the phonon and ionization channel surfaces and 0.1\% for the unpatterned sidewalls.

Holding all other quantities fixed, we varied the amount of phonon energy lost from the crystal (absorbed into the readout channels) when a phonon interacts in the aluminum $\mathcal{L}$ from values of 2, 3, 4, 6 and 8 times the aluminum superconducting gap $E_g$. This model is a proxy for a more complicated model involving quasiparticle downconversion in the aluminum and 2~$\times E_g$ is the minimum amount of energy that must be removed in a phonon interaction. The removal of 4~$\times E_g$ is similar to the more complete Kaplan downconversion model when the aluminum film thickness of of 300~nm and an interaction length of 720~nm is considered. We also considered variations in the TES superconducting temperature $T_c$ of 50, 55 and 60~mK. Finally we investigated possible contributions to the pulse shape from phonon physics and varied the anharmonic decay constant $A$ from 1.61$\times 10^{-55}$, 9.05$\times 10^{-56}$, 5.37$\times 10^{-56}$, 2.86$\times 10^{-56}$ and 1.61$\times 10^{-56}$; where the first value 1.61$\times 10^{-55}$ is found in the literature and would provide the quickest rise times. 

\begin{figure}[t]
\begin{tabular}{c}
\begin{minipage}{1\hsize}
\begin{center}
\includegraphics[width=12cm]{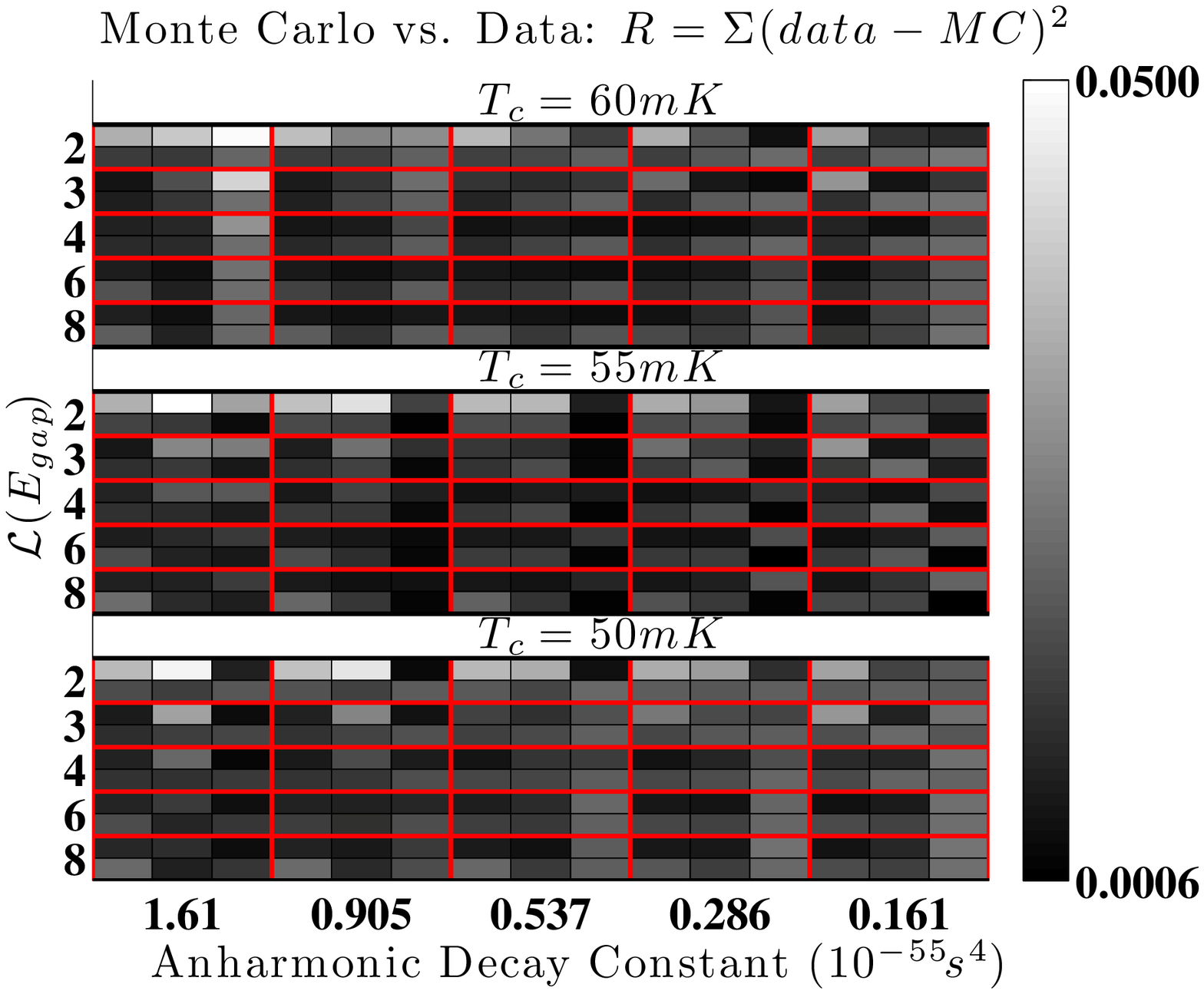}
\end{center}
\end{minipage}
\\
\begin{minipage}{1\hsize}
\begin{center}
\includegraphics[width=15cm]{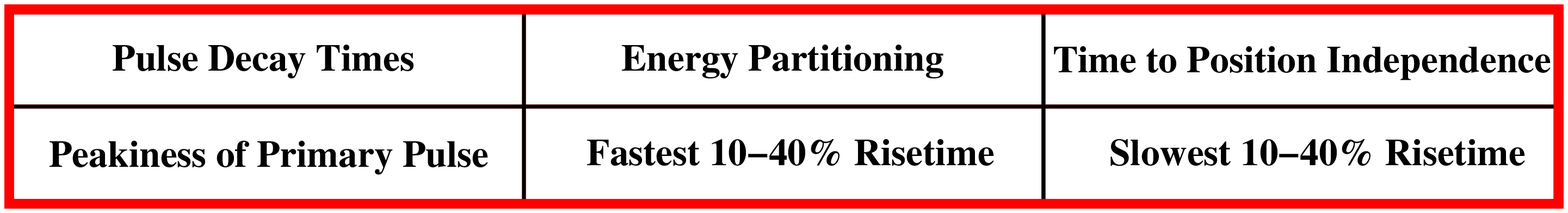}
\end{center}
\end{minipage}
\end{tabular}
\caption[] { \label{fig:allchi2} Likelihood values for agreement between data and Monte Carlo histograms for various pulse shape quantities. Each rectangle bounded by red lines corresponds to a distinct $T_c, \mathcal{L}$ and~$A$ and contains 6 patches that indicate the R-values of 6 distinct pulse shape parameters (top row l-r) decay time, energy patitioning, time to position independence (bottom row, l-r) peak height to energy ratio, fastest risetime, slowest risetime (as shown in the legend). The y-axis, which corresponds to $\mathcal{L}$, is repeated three times, once for each of the 3 $T_c$ values. R-values for a pulse shape parameter are normalized so that the sum of the 75 P-values for each parameter equals 1.}
\end{figure}

\begin{figure}[t]
\begin{center}
\includegraphics[width=12cm]{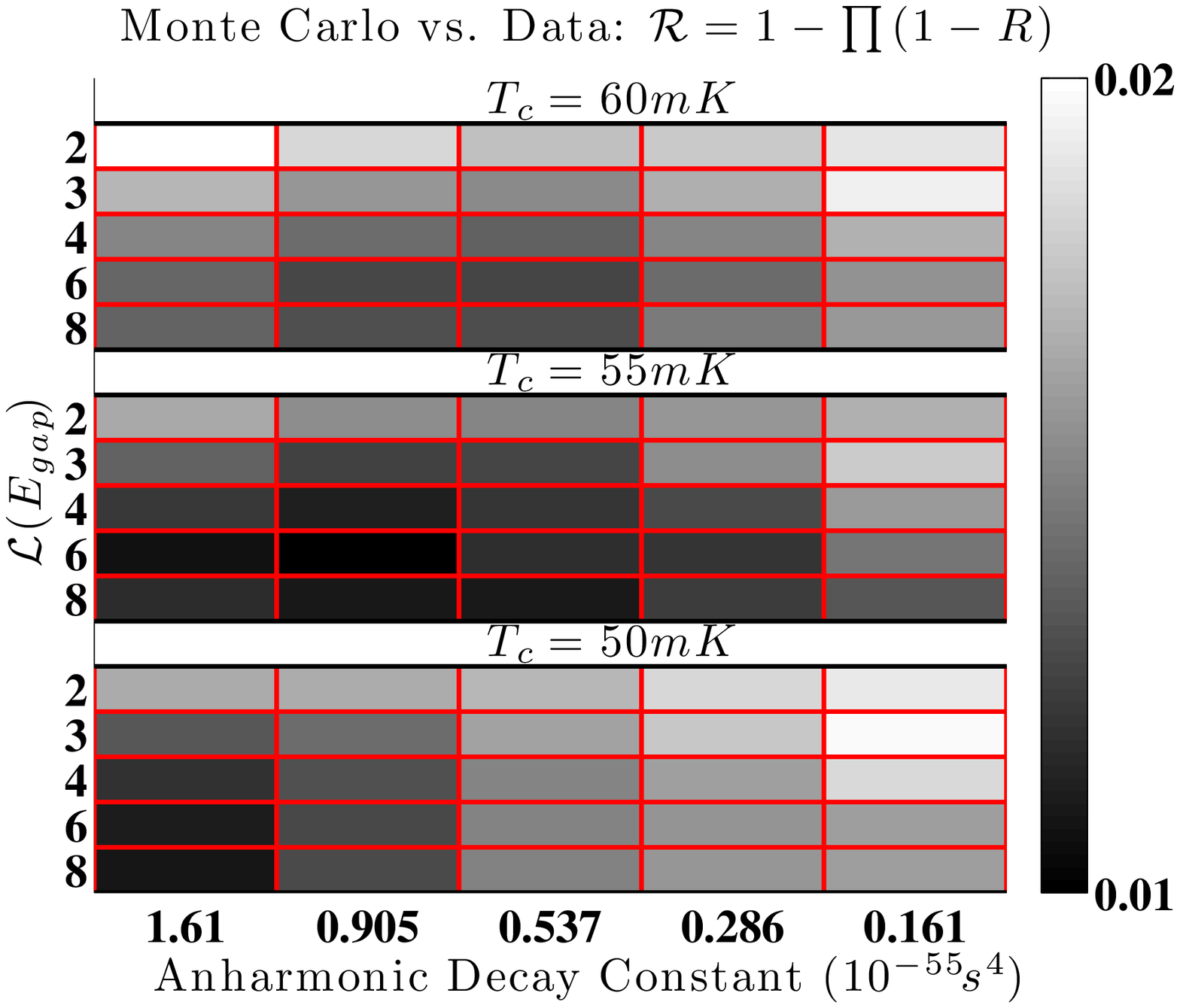}
\end{center}
\caption[] { \label{fig:combinedchi2} Combined pulse shape likelihood parameters $\mathcal{R}$. }
\end{figure}

Various pulse shape parameters are considered in comparing calibration data and Monte Carlo: the fraction of energy deposited in each channel, pulse decay times, pulse risetimes, peak time, peak height to energy ratio and time to loss of position sensitivity. A least-squares residual ($R=\sum{(\mathit{data}-\mathit{MC})^2}$, where the data and Monte Carlo distributions are normalized to area = 1) likelihood map of these parameters is shown in Figure~\ref{fig:allchi2}. A combined likelihood ($\mathcal{R} = 1-\prod{(1-R)}$, where each of the six residuals $R$ are normalized to sum to unity) combining these six pulse shape parameters is shown in Figure~\ref{fig:combinedchi2}. While there are unaccounted for correlations in the pulse shape parameters that went into this joint likelihood it nonetheless is in agreement with our expectations. Given the coarse 5~mK sampling in T$_c$, there is still the possibility of some pulse shape discrepancy between Monte Carlo and data. Given a prior belief in a $1.61\time10^{-55}$ anharmonic decay constant and a mean $4 \times E_g$ energy loss in a phonon-aluminum interaction, a $T_c~<$~55~mK is favored. Various distributions are discussed in more detail along with influences of $A$, $T_c$ and $\mathcal{L}$.

\begin{figure}
\begin{tabular}{cc}

\begin{minipage}{0.48\hsize}
\begin{center}
\includegraphics[width=7cm]{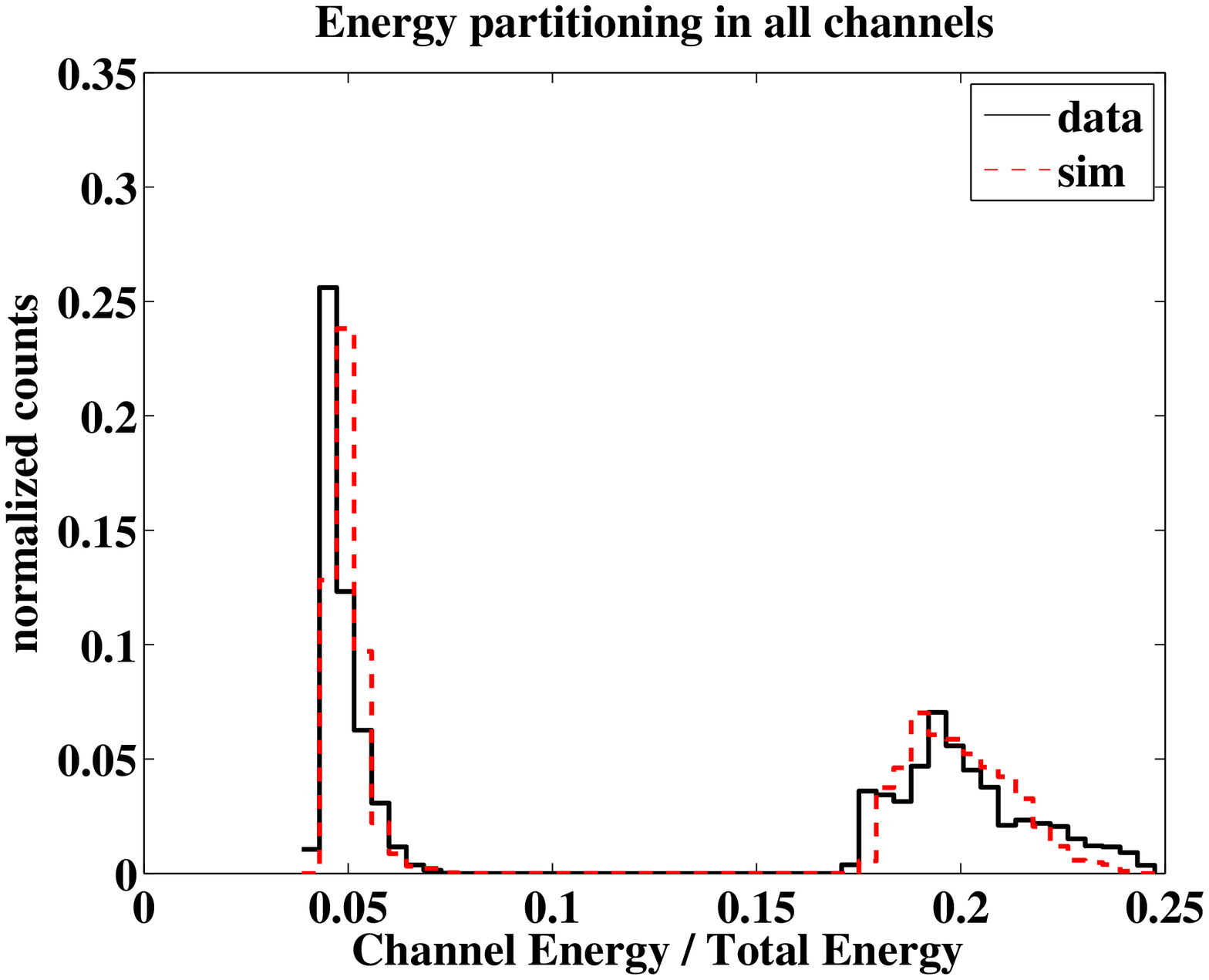}
\parbox{6cm}{\caption[] { \label{fig:Energy_hists} Distribution of energy deposited in each phonon channel. The outer channels are of reduced area coverage and contribute to the distribution at 0.05.}}
\end{center}
\end{minipage}

\begin{minipage}{0.48\hsize}
\begin{center}
\includegraphics[width=7cm]{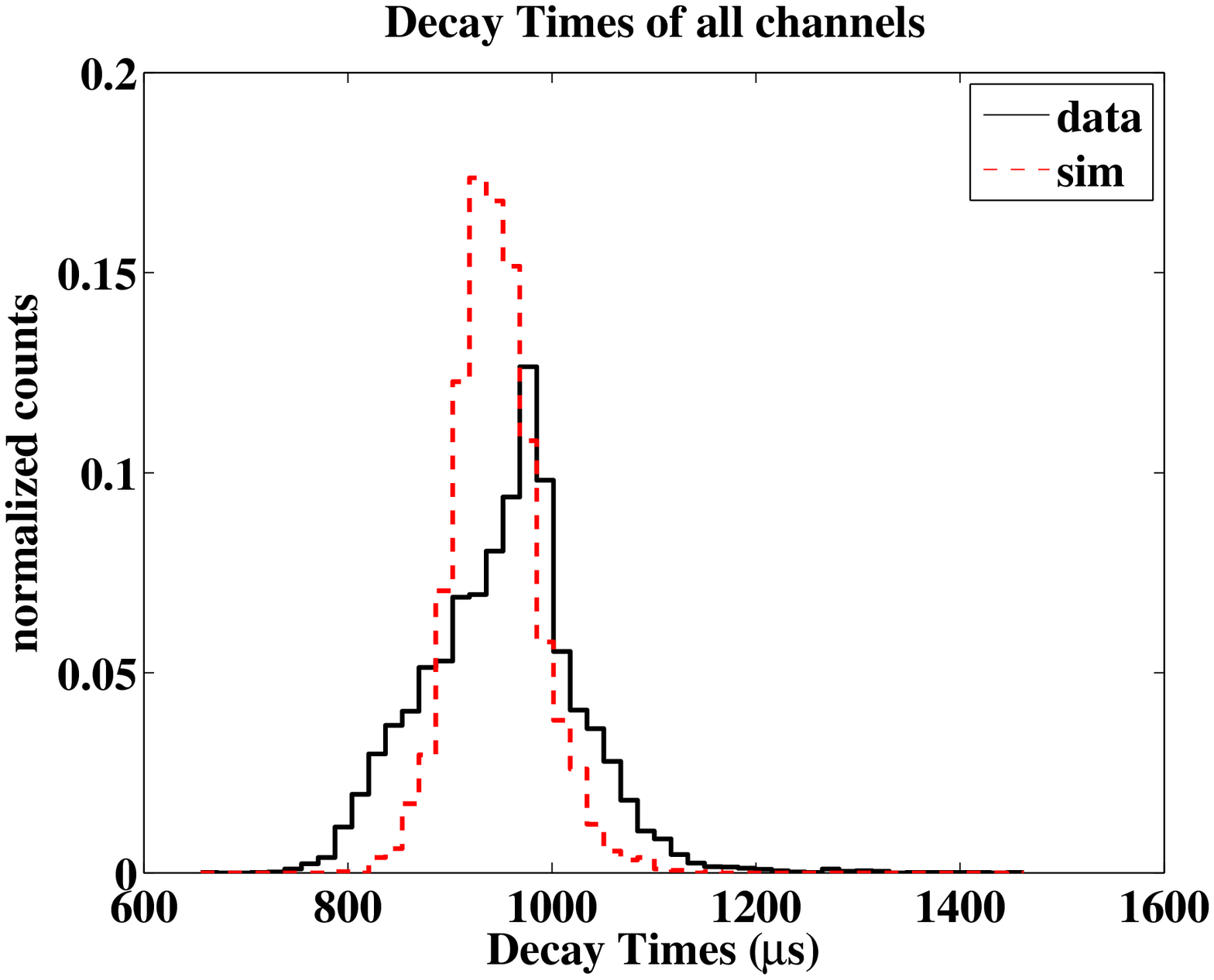}
\parbox{6cm}{\caption[] { \label{fig:Decay_hists} Distribution of pulse decay times. The data shows systematic differences between channels and leads to large spread when combined.}}
\end{center}
\end{minipage}

\end{tabular}
\end{figure}

The fraction of energy absorbed in each channel is shown in figure~\ref{fig:Energy_hists}. The channels at high radius are of smaller area coverage and thus absorb a smaller fraction of the total phonon energy; they populate the distribution at 0.05. The inner channels are larger and populate the distribution at 0.2. This region is multi peaked reflecting events that occur near (away from) the outer channels with reduced (increased) energy. This distribution is strongly affected by the phonon loss value $\mathcal{L}$ and mildly by $T_c$. Channels with large energy deposition result in larger TES temperature excursions and hence more energy re-emitted into the crystal. This energy is presumably below the aluminum superconducting gap and not detected by another channel.

The decay times are shown in figure~\ref{fig:Decay_hists}. Four of the channels in the detector show tight distributions, similar to that in Monte Carlo while four others have more spread. The agreement in distribution centers shows that the DMC has the phonon-aluminum interaction probability and $\mathcal{L}$ values reasonably set. The width of the data's distribution is not completely understood and may be due to phase separation in the TES system, something that is being more extensively studied in future detectors.

Obtaining good pulse shape parameters in the early portion of the pulse presents more of a challenge. The actual physical processes controlling these features are more complicated and rely on more precise tuning of Monte Carlo models including quasiparticle phonon downconversion, Luke phonons and perhaps most importantly, TES dynamics and phase separation.

\begin{figure}
\begin{tabular}{cc}

\begin{minipage}{0.48\hsize}
\begin{center}
\includegraphics[width=7cm]{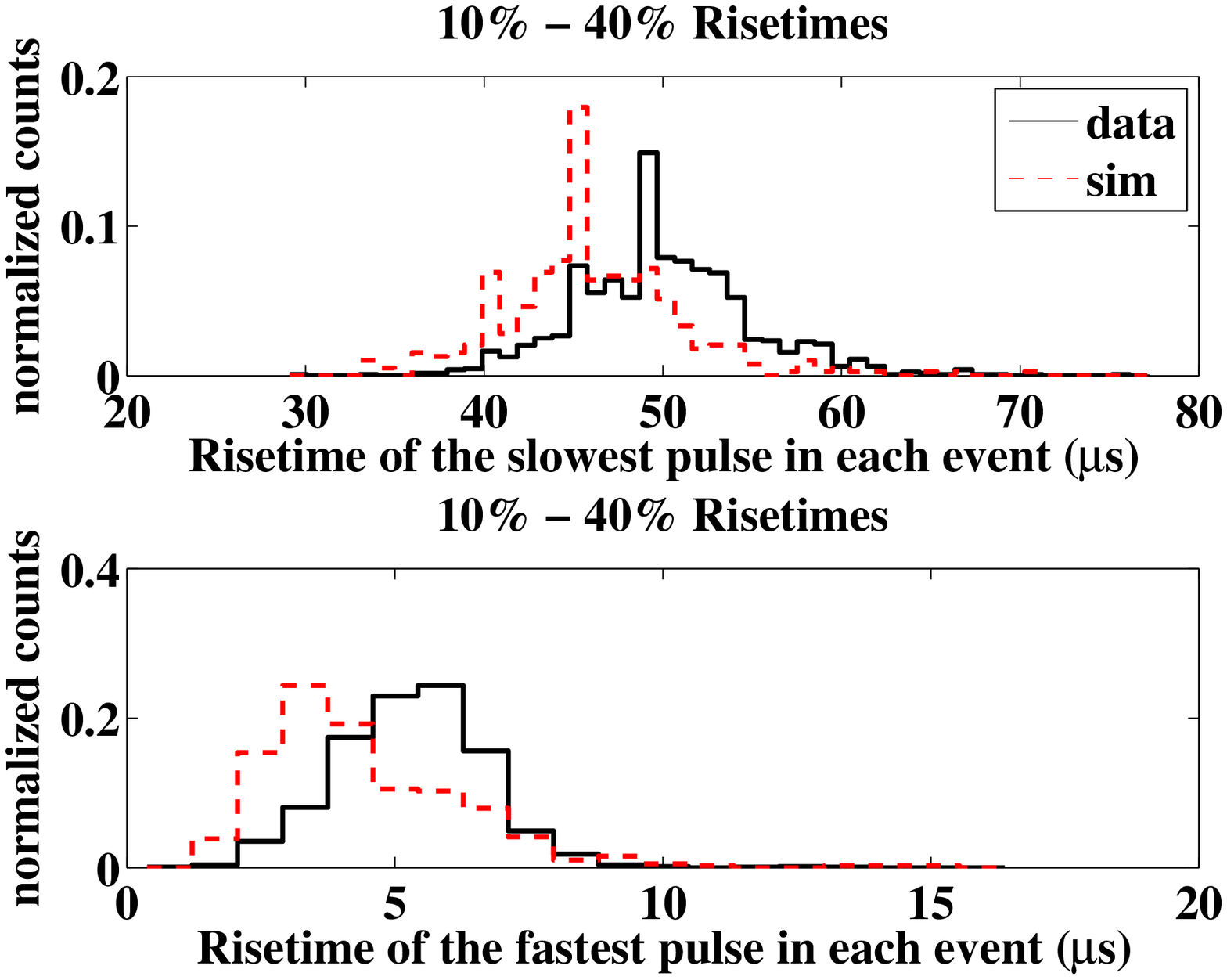}
\parbox{6cm}{\caption[] { \label{fig:RT1040_hists} Distribution of 10-40\% risetimes for the slowest and fastest channels.}}
\end{center}
\end{minipage}

\begin{minipage}{0.48\hsize}
\begin{center}
\includegraphics[width=7cm]{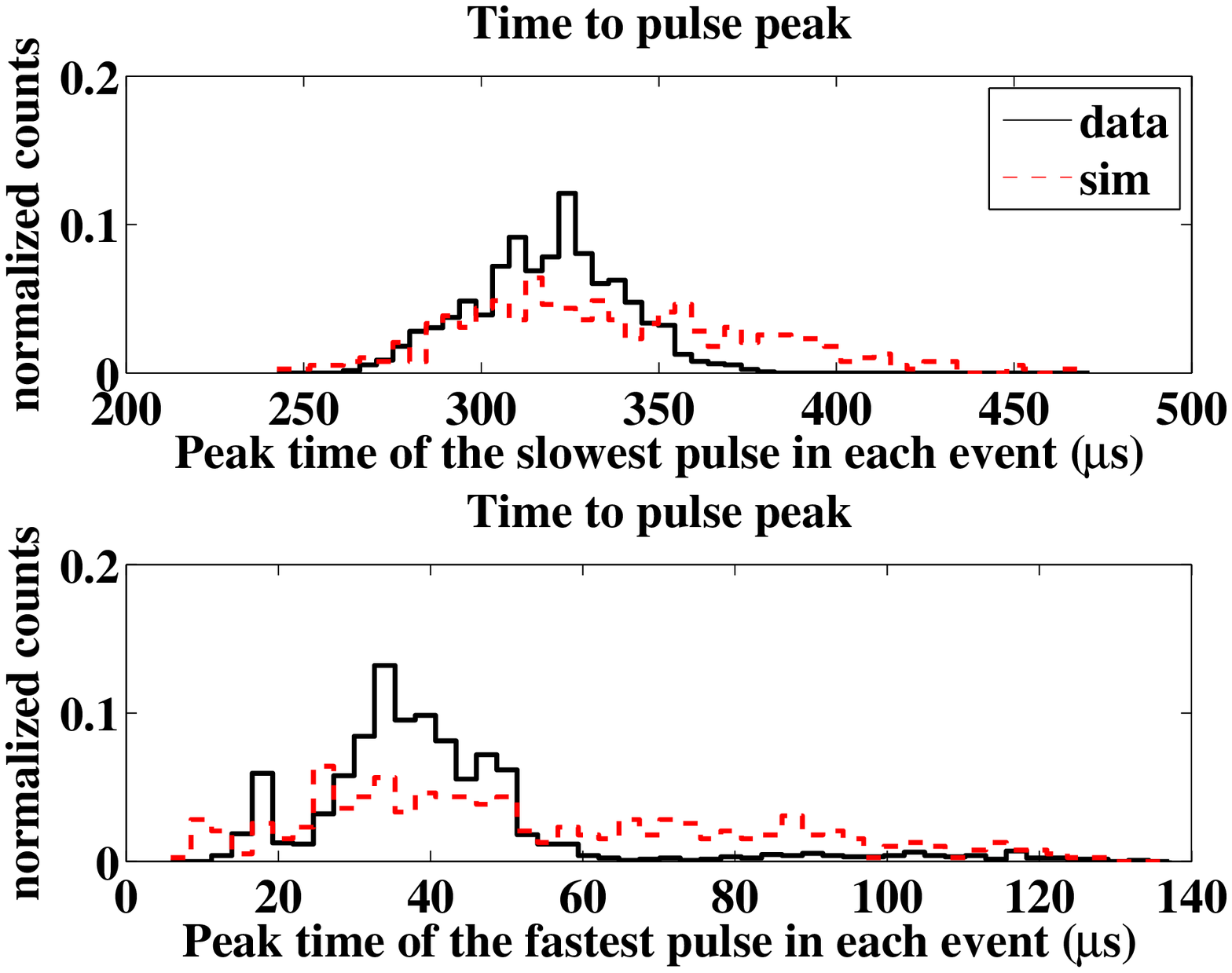}
\parbox{6cm}{\caption[] { \label{fig:Peaktime_hists} Peak time distributions for the fastest and slowest channels.}}
\end{center}
\end{minipage}

\\

\begin{minipage}{0.48\hsize}
\begin{center}
\includegraphics[width=7cm]{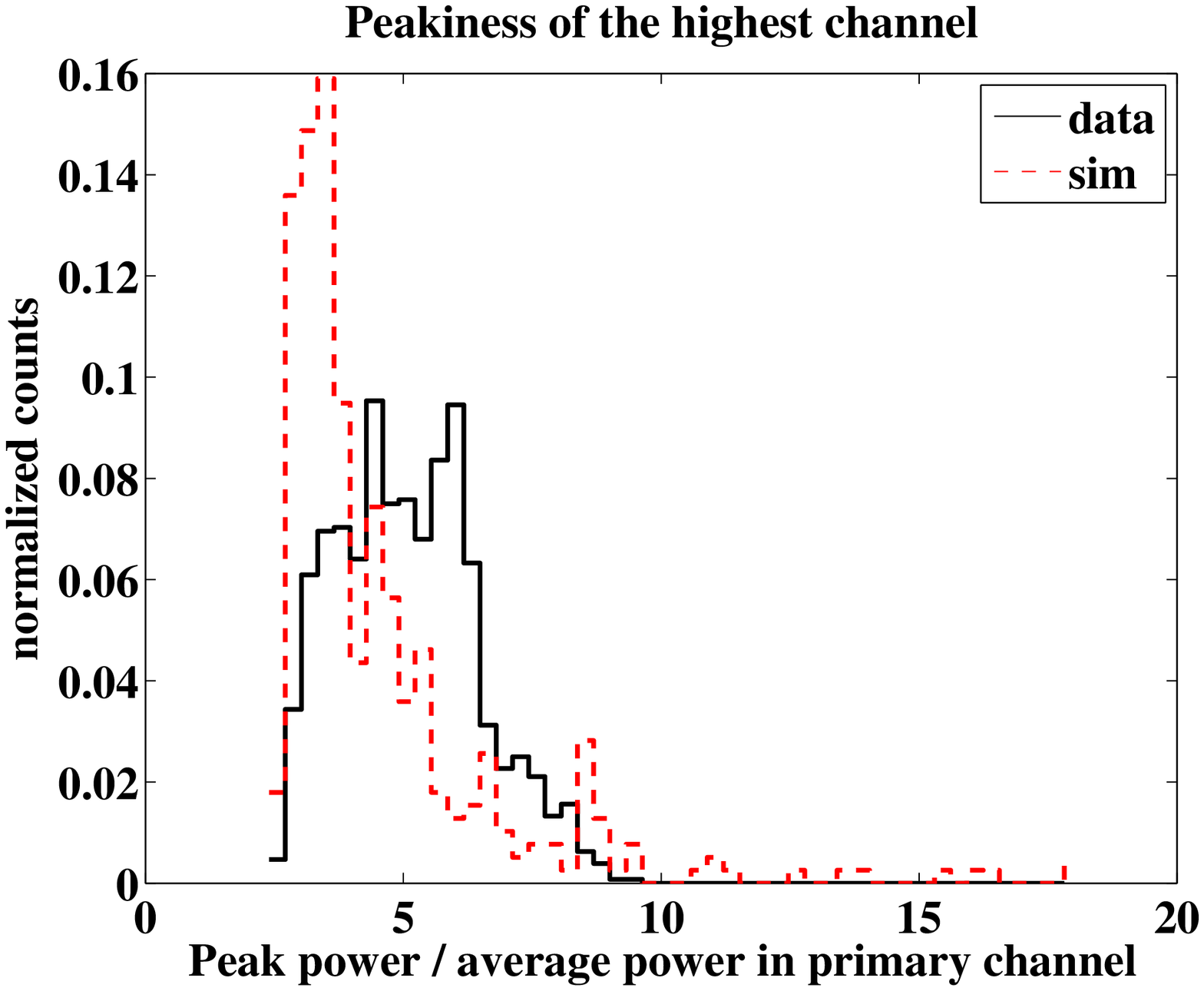}
\parbox{6cm}{\caption[] { \label{fig:Peakheight_hists} Distributions of pulse ``peakiness'', pulse height to energy ratio.}}
\end{center}
\end{minipage}

\begin{minipage}{0.48\hsize}
\begin{center}
\includegraphics[width=7cm]{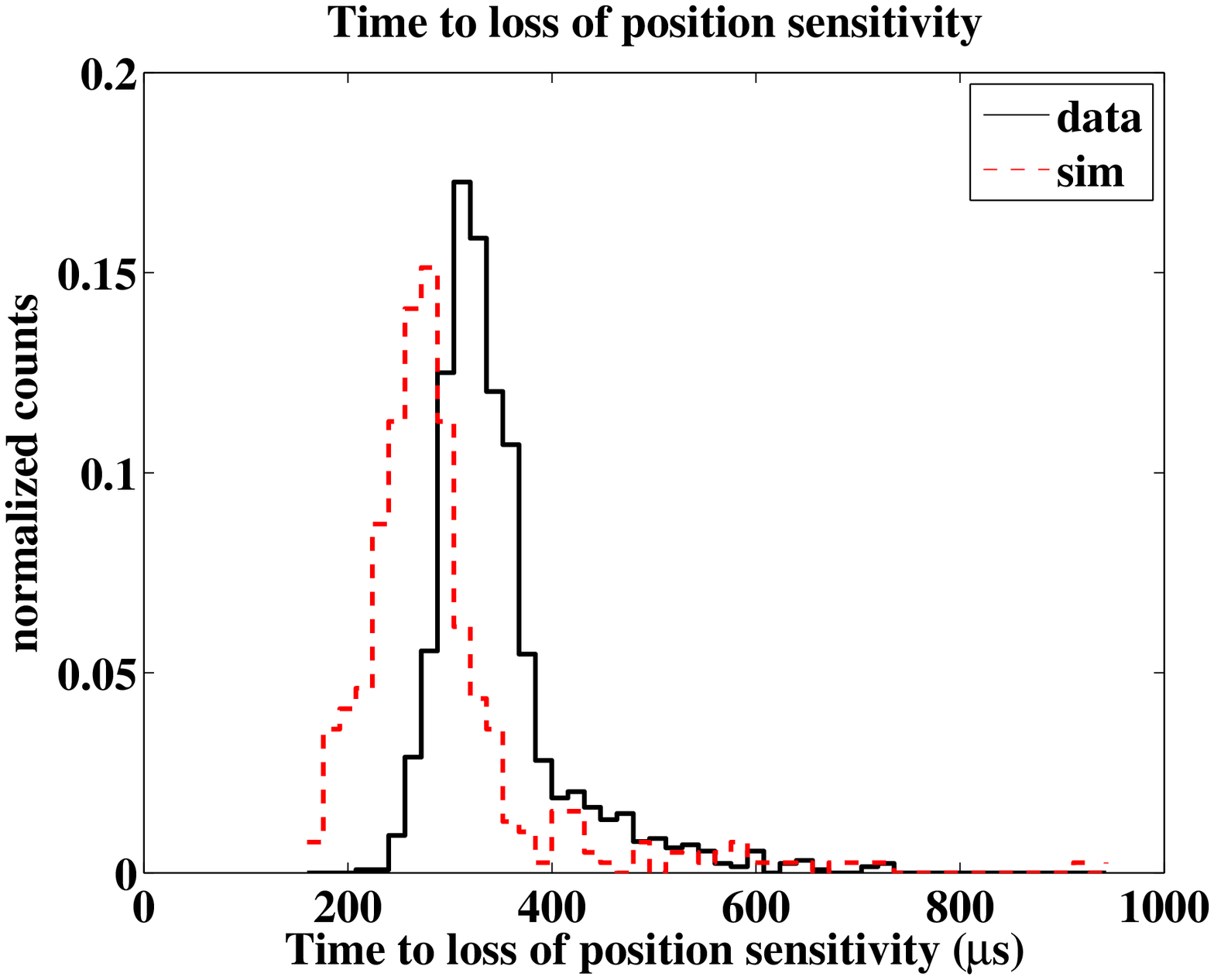}
\parbox{6cm}{\caption[] { \label{fig:Pos_Ind_Time_Hist_wdata} Time to loss of position sensitivity in pulse decay.}}
\end{center}
\end{minipage}

\end{tabular}
\end{figure}

The distribution of 10-40\% risetimes, for both the slowest and fastest channels is shown in figure~\ref{fig:RT1040_hists}. The fastest channel risetimes show overlapping distributions whereas the slowest channels are consistent within ~10\%. This is good agreement for a challenging parameter and would likely be improved upon by running Monte Carlo over additional $T_c$.

The peak time distributions for the slowest  and fastest channels as shown in figure~\ref{fig:Peaktime_hists}. The slowest channel peak times are well reproduced in Monte Carlo indicating that the anharmonic decay constant $A$ is consistent with the published value. The same distributions show poorer agreement however when the faster channels are considered, again a challenging test in Monte Carlo. The peak height to energy ratio for the primary channels is shown in figure~\ref{fig:Peakheight_hists}. Combined, these could continue to indicate that $T_c$ and $\mathcal{L}$ are not sufficiently tuned or that phase separation is at play.

Another important metric in our tuning was the time until position information was lost in the pulses, when the pulses reached a mean decay. We defined this as the time when all traces reached within 10\% of the mean. The distribution is shown in Figure~\ref{fig:Pos_Ind_Time_Hist_wdata} and shows reasonably good agreement between calibration data and Monte Carlo. This parameter is degenerate in $A$ and $T_c$ and relatively insensitive to $\mathcal{L}$.

\section{Conclusion}

The phonon quasidiffusion measurements presented here occurred on large 3~inch diameter, 1~inch thick high purity germanium crystals. These crystals were cooled to 50~mK in the vacuum of a dilution refrigerator. Good agreement is seen between Am-241 calibration data and a detailed Monte Carlo for anharmonic decay constant $A = 1.61\time10^{-55}$, a mean phonon-aluminum interaction loss $\ge 4 \times E_g$ and a TES $T_c~<$~55~mK. Future work will involve testing on detectors with $T_c$ that is more easily measured, including electric field models into the Monte Carlo which include radial fringing fields and a detailed Kaplan quasiparticle-phonon downconversion model.

\section{Acknowledgements}

This research was funded in part by the Department of Energy (Grant Nos. DE-FG02-04ER41295 and DE- FG02-07ER41480) and by the National Science Foundation (Grant Nos. PHY-0542066, PHY-0503729, PHY-0503629, PHY-0504224, PHY-0705078, PHY-0801712)

\bibliography{aipsamp}% Produces the bibliography via BibTeX.

\clearpage


\section{References}





\begin{thebibliography}{}

\bibitem{Ahmed2009} Z. Ahmed, Phy. Rev. Lett. {\bf 102}, 011301 (2009)

\bibitem{Spergel2007} D. N. Spergel et al. (WMAP Collaboration), Astrophys. J. Suppl. Ser. {\bf 170}, 377 (2007)

\bibitem{Tegmark2004} M. Tegmark et al. (SDSS Collab.), Phys. Rev. D {\bf 69}, 103501 (2004)

\bibitem{Neganov1985} B. Neganov and V. Trofimov, USSR Patent No 1037771, Otkrytia i izobreteniya {\bf 146}, 215 (1985)
        
\bibitem{Luke1988} P.N. Luke, J. Appl Phys. {\bf 64}, 12 (1988)

\bibitem{Marris1990} H.J. Marris, Phys. Rev. B {\bf 41}, 9736 (1990)

\bibitem{Tamura1993} S. Tamura, Phys. Rev. B {\bf 48}, 13502 (1993)

\bibitem{Irwin1995} K.D. Irwin et al., IEEE Trans on App Supercond., vol 5, 2690 (1995)

\bibitem{Msall1997} M.E. Msall et al, Phys. Rev. B, {\bf 56}, 9557 (1997)



\bibitem{Oed1988} A. Oed, Nuclear Instruments and Methods in Physics Research, A263 (1988) 351-359

\bibitem{Luke1994} P.N. Luke, Applied Physics Letters {\bf 65} (22) 1994

\bibitem{Brink2006} P.L. Brink et al, Nuclear Instruments and Methods in Physics Research, A559 (2006) 414Ð416

\bibitem{Leman2011_5} S.W. Leman, Review of Physics and Monte Carlo Techniques as Relevant to a Cryogenic, Phonon and Ionization Readout, Radiation-Detector, submitted 2011, Review of Modern Physics


\bibitem{Yellin2002} S. Yellin, Phys. Rev. D {\bf 66}, 032005 (2002)

\bibitem{Kittel1980} C. Kittel and H. Kroemer, Thermal Physics (W.H. Freeman and Company, 1980)

\bibitem{Wolfe1998} J.P. Wolfe, Imaging Phonons: Acoustic Wave Propagation in Solids (Cambridge U. Press, 1998)

\bibitem{Tamura1993_2} S. Tamura, Phys. Rev. B {\bf 48}, 13502 (1993)

\bibitem{Sasaki1958} W. Sasaki et al, Journal of the Physical Society of Japan, {\bf 13}, 456 (1958)

\bibitem{Jacoboni1983} C. Jacoboni et al, Rev. Mod. Phys., {\bf 55}, 645 (1983)

\bibitem{Cabrera} B. Cabrera, submitted to Applied Physics Letters 

\bibitem{Herring1956} C. Herring et al, Phys. Rev, {\bf 101}, 944 (1956)

\bibitem{Sundqvist2009} K.M. Sundqvist, AIP Conference Series, {\bf 1185}, 128 (2009)

\end{thebibliography}

\end{document}